\documentclass[conference]{IEEEtran}

\ifCLASSINFOpdf
 
\else
 
\fi

\usepackage{array}
\usepackage{amsthm,amsmath,amssymb}
\usepackage{mathrsfs}
\usepackage{multirow}
\usepackage{siunitx}

\usepackage{threeparttable}
\usepackage{booktabs}
\usepackage{enumerate}
\usepackage{cite}
\usepackage{graphicx}
\usepackage[table]{xcolor} 
\usepackage{colortbl}
\usepackage{notation}
\usepackage{soul}

\begin{document}
\newtheorem{Def}{Definition}
\newtheorem{Pro}{Proposition}
\newtheorem{Lem}{Lemma}
\newtheorem{Exa}{Example}
\newtheorem{Rem}{Remark}
\newtheorem{Cor}{Corollary}
\renewcommand{\labelitemi}{$\bullet$}

\newcommand{\ist}{\hspace*{.3mm}}
\newcommand{\rmv}{\hspace*{-.3mm}}
\newcommand{\iist}{\hspace*{1mm}}
\newcommand{\rrmv}{\hspace*{-1mm}}
\newcommand{\nn}{\nonumber}
\newcommand{\sist}{\hspace*{.15mm}}
\newcommand{\trans}{^\mathrm{T}}

\title{A New Architecture for\\ Neural Enhanced Multiobject Tracking}
\author{\IEEEauthorblockN{Shaoxiu Wei, Mingchao Liang, and Florian Meyer}\vspace{.8mm}
\IEEEauthorblockA{University of California San Diego, La Jolla, CA (\{shwei,m3liang,flmeyer\}@ucsd.edu)\vspace{1.5mm}}
 \vspace*{-7mm}}

\providecommand{\bu}{\textcolor{blue}}
\providecommand{\rd}{\textcolor{red}}
\maketitle

\IEEEpeerreviewmaketitle

\begin{abstract}
	Multiobject tracking (MOT) is an important task in robotics, autonomous driving, and maritime surveillance. Traditional work on MOT is model-based and aims to establish algorithms in the framework of sequential Bayesian estimation. More recent methods are fully data-driven and rely on the training of neural networks. The two approaches have demonstrated advantages in certain scenarios. In particular, in problems where plenty of labeled data for the training of neural networks is available, data-driven MOT tends to have advantages compared to traditional methods. A natural thought is whether a general and efficient framework can integrate the two approaches. This paper advances a recently introduced hybrid model-based and data-driven method called neural-enhanced belief propagation (NEBP). Compared to existing work on NEBP for MOT, it introduces a novel neural architecture that can improve data association and new object initialization, two critical aspects of MOT. The proposed tracking method is leading the nuScenes LiDAR-only tracking challenge at the time of submission of this paper\vspace{1mm}.
\end{abstract}
\IEEEpeerreviewmaketitle

\section{Introduction}
Multiobject tracking (MOT)\cite{MTT, RadarTrackingBook-1986Blackman}  is an important capability in a variety of applications, including autonomous navigation, applied ocean sciences, and aerospace surveillance. In MOT, the number of objects to be tracked is unknown, and there is measurement-origin uncertainty, i.e., it is unknown which object generated which measurements. Thus, key aspects of MOT methods are data association, object track initialization, and sequential estimation. Traditional model-based MOT methods have evolved from Bayesian filtering theory and perform object track initialization, motion prediction, data association, and measurement updates based on well-established statistical models. Model-based MOT approaches include vector-type methods such as the joint probabilistic data association (JPDA) filter\cite{JPDA-1998}, the multiple hypothesis tracker (MHT)\cite{MHT-2004Blackman}, and belief propagation (BP) \cite{BP-2018Meyer}. Recent developments in this Bayesian framework\cite{BayesianFilteringBook-2013} also include set-type methods\cite{RFSbook-2014Mahler} such as the probability hypothesis density (PHD) filter \cite{PHD-2006Vo}, the labeled multi-Bernoulli (LMB) filter \cite{LMB-2014Reuter}, and Poisson multi-Bernoulli (PMB) filters \cite{Wil:J15,BP-2018Meyer,PMBM-2018Angel}. To obtain tractable algorithms, model-based methods are typically implemented using a Gaussian mixture or particle-based representation of probability density functions (PDFs). Model-based MOT methods have been tailored to specific applications by introducing additional parameters to be estimated and tracked, such as the time-varying detection probabilities \cite{UnknownParameter-2019Meyer, TvT-Shaoxiu}, maneuvering motion\cite{UnknownParameter-2019Meyer,Manoeuvring-2020Godsill}, and objects extents\cite{ExTrackOverview-2017Granstrom,ExTrack-2021Meyer}.


A recent trend in MOT is using data-driven methods that make use of neural architectures such as the multi-layer perceptrons (MLPs) \cite{Shasta-2023sadjadpour}, graph neural networks (GNNs)\cite{GNN3DMOT-2020Weng,GNNMOT-2020Guillem,GNNMOT-2022Martin,GNNonline-2022Zaech}, and transformers\cite{FormerMOT3D-2023Shuxiao, TransformerbasedMOT-2023pinto, TrackFormer-2022Tim} that are trained with labeled data. Although their structures vary, the primary objective of most data-driven methods is to extract information for data association and object track estimation. For example, \cite{GNNMOT-2020Guillem} used a GNN to perform data association. Due to the expressive power of neural networks\cite{Voxelnet-2018Zhou}, data-driven methods are able to extract complex features from data that are difficult to be described by a statistical model.
\begin{figure}[!t]
	\centering
	\includegraphics[width=3.4in]{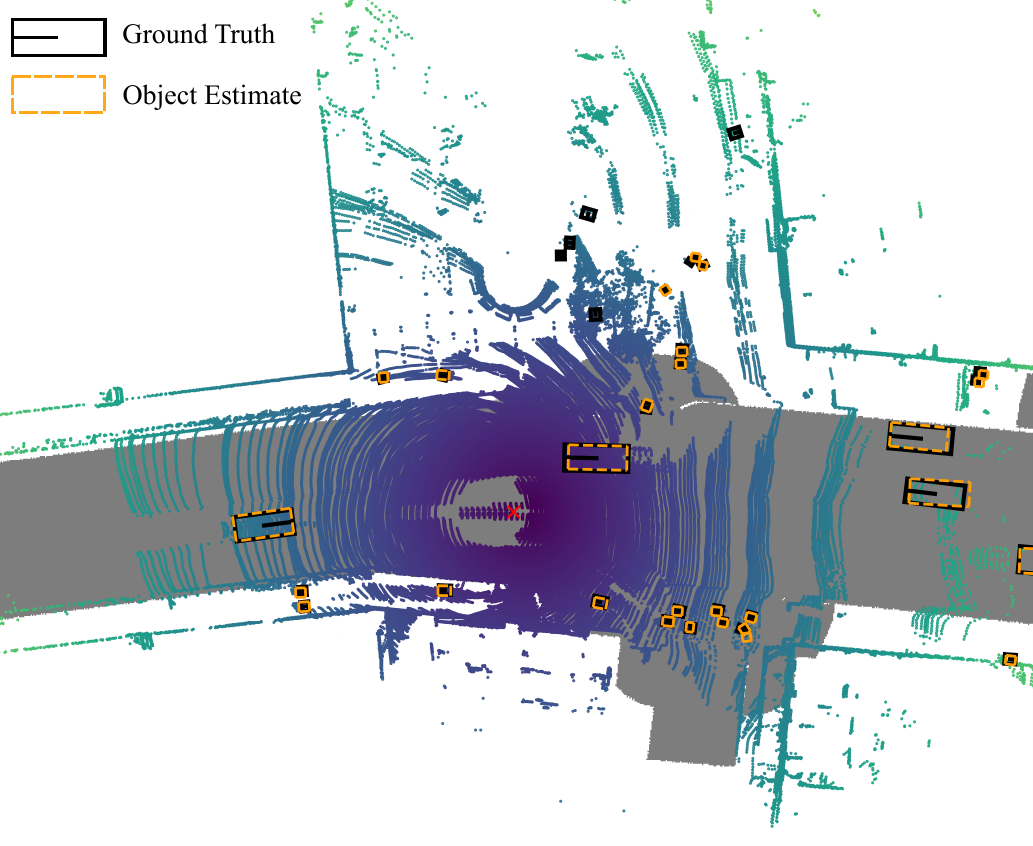}
	\caption{Considerd multiobject tracking scenario. LiDAR measurements, ground truth, and tracking results of the proposed method. The orange dashed rectangles indicate the object estimates and the black rectangles indicate ground truth.\vspace{-5.5mm}.}
	\label{Nuscenes}
\end{figure}

Recently, neural-enhanced belief propagation (NEBP) for MOT was introduced to combine the advantages of model-based and data-driven approaches  \cite{LiaMey:J24}. NEBP combines BP-based MOT \cite{MeyBraWilHla:J17,BP-2018Meyer} with a GNN that extracts features from raw sensor data and exchanges information with model-based BP. One significant advantage compared to fully data-driven methods is that established statistical models can still be utilized. At the same time, neural networks enhance traditional models with information learned from raw sensor data. The original NEBP approach can achieve state-of-the-art performance in autonomous driving scenarios \cite{LiaMey:J24}. In this paper, we propose NEBP+ for MOT. Compared to the original NEBP approach \cite{LiaMey:J24}, NEBP+ introduces an improved neural architecture and makes use of additional types of features. The key insights leveraged by the neural architecture of NEBP+ is that the overall MOT performance can be increased significantly by (i) using precomputed differences of features as input to the neural architecture and (ii) fusing different feature similarities based on learnable weights. The contributions of this paper can be summarized as follows\vspace{0.5mm}.

\begin{enumerate}		
	\item We propose a novel neural network architecture that enhances the messages of BP-based MOT by learned information and features extracted from raw sensor data\vspace{0.8mm}.
	\item The proposed NEBP+ method is applied to the nuScenes autonomous driving dataset\cite{Nuscenes} and can achieve state-of-the-art object tracking performance\vspace{0.8mm}.
\end{enumerate}
At the time of submission of this paper, the proposed tracking method is leading the nuScenes LiDAR-only tracking challenge \cite{Nuscenes}\vspace{2mm}.

\section{System Model and Problem Formulation}
In this section, we will introduce the basic assumptions for Bayesian MOT and establish the used notation.
\subsection{Object States and Transition Model}
\par At time $k$, there are $N_k$ potential objects (POs) $\V{x}^i_{k}$, $i\in\{1,2,\dots,N_k\}$, where $N_k$ is the maximum possible number of objects that have generated a measurement, and the existence of each PO is modeled by a binary random variable $r^i_{k}\in \{0, 1\}$. A PO $\V{x}^i_{k}$ exists if and only if $r^i_{k}=1$. For simplicity, we use the notation ${\V{y}}^i_{k} = [(\V{x}^i_{k})^\top~r^i_{k}]^{\top}$ for a PO state that has been augmented by an existence variable. An object detector is applied to sensor data and produce $J_k$ measurements denoted as $\M{z}_k = [({\V{z}^1_{k}})^{\top}~({\V{z}^2_{k}})^{\top}\cdots({\V{z}^{J_k}_{k}})^{\top}]^{\top}\rmv$.
\par Generally, there are two types of POs:
\begin{enumerate}		
	\item \emph{The Legacy POs}: $\underline{\V{y}}^i_{k} = [(\underline{\V{x}}^i_{k})^{\top}~\underline{r}^i_{k}]^{\top}, i \in\{1,2,\dots,I_k\}$ represent objects that have generated a measurement at a previous time step $k' < k$. The joint state of legacy POs is defined as $\underline{\V{y}}_{k} = [(\underline{\V{y}}^1_{k})^\top\cdots(\underline{\V{y}}^{I_k}_{k})^\top]^\top\rmv\rmv$\vspace{.5mm}.
	\item \emph{The New POs}: $\overline{\V{y}}^j_{k} = [(\overline{\V{x}}^j_{k})^\top~\overline{r}^j_{k}]^{\top}, j \in\{1,2,\dots,J_k\}$ represent objects that generate a measurement at time $k$ for the first time. Each measurement $\V{z}^j_{k}, j \in \{1,2,\dots,J_k\}$ will form a new PO $\overline{\V{y}}^j_{k}$. The joint state of new POs is defined as $\overline{\V{y}}_{k} = [(\overline{\V{y}}^1_{k})^\top\cdots(\overline{\V{y}}^{J_k}_{k})^\top]^\top\rmv\rmv$.
\end{enumerate}
When the measurements of the next time step are considered, new POs will become legacy POs. Thus, the number of legacy POs at time $k$ is $I_k = I_{k-1}+ J_{k-1}$ and the total number of POs is $N_k = I_k + J_k$.  The joint PO states $\V{y}_k$ at time $k$ is defined as $\V{y}_k = [(\underline{\V{y}}_{k})^\top~(\overline{\V{y}}_{k})^\top]^\top$.
\par Given the PO state ${\V{y}}^i_{k-1}$ at time $k-1$, its transition to the legacy PO ${\underline{\V{y}}}^i_{k}$ at time $k$ is assumed to be independent and identical according to a Markovian dynamic model\cite{MTT}. Then, the transition PDF of the joint PO states $\V{y}_{k-1}$ at time $k-1$ can be expressed as
\begin{align}
	f({\underline{\V{y}}}_{k}|&{\V{y}}_{k-1}) =\!\!\! \prod_{i=1}^{N_{k-1}}\!\!{f({\underline{\V{x}}}^i_{k},\underline{r}^i_{k}|{\V{x}}^i_{k-1}, {r}^i_{k-1})}. \nn
\end{align}
If the PO $i$ does not exist at time $k-1$, i.e., $r_{k-1}^i = 0$, it cannot exist at time $k$. Thus, the state transition PDF can be expressed\vspace{-1.5mm} as
\begin{align} 
	f({\underline{\V{x}}}^i_{k},\underline{r}^i_{k}|{\V{x}}^i_{k-1}, 0)  =	\left\{
	\begin{array}{lr}
		0, &\underline{r}^i_{k} = 1\\
		f_{D}(\underline{\V{x}}^i_{k}), &\underline{r}^i_{k} = 0
	\end{array}
	\right. \nn
\end{align}
where $f_{D}(\underline{\V{x}}^i_{k})$ is an arbitrary ``dummy'' PDF since states of nonexisting POs are irrelevant. If the PO $i$ exists at time $k-1$, i.e. $r_{k-1}^i = 1$, then it continues to exist with a survival probability $p_s$\vspace{-1mm}, i.e.,
\begin{align} \nn
	f({\underline{\V{x}}}^i_{k},\underline{r}^i_{k}|{\V{x}}^i_{k-1}, 1)  =	
	\left\{
	\begin{array}{lr}
		p_s f({\underline{\V{x}}}^i_{k}|{\V{x}}^i_{k-1}), &\underline{r}^i_{k} = 1 \\
		(1 - p_s)f_{D}(\underline{\V{x}}^i_{k}), &\hspace{1mm} \underline{r}^i_{k} = 0.
	\end{array}
	\right. 
\end{align}

\subsection{Data Association and Measurement Model}
\par In this paper, we only consider the point object tracking, i.e., each object can generate at most one measurement, and each measurement can originate from at most one object. The former is described by the ``object-oriented'' data association vector $\V{a}_k = [a_k^1~a_k^2~{\cdots}~a_k^{I_k}]^\top$\rmv\rmv\rmv, where each element takes value in $\{0,1,2,\dots,J_k\}$. The latter is described by the ``measurement-oriented'' data association vector $\V{b}_k = [b_k^1~b_k^2~\cdots~b_k^{J_k}]^\top$\rmv\rmv\rmv, where each element takes value in $\{0,1,2,\dots,I_k\}$. Under the data association assumption, they are equivalent since one can be determined from the
other \cite{BP-2018Meyer}. The indicator function is used to check whether $\V{a}_k$ and $\V{b}_k$ are consistent,\vspace{-.5mm} i.e.,
\begin{align}
	\Phi(\V{a}_k,\V{b}_k) = \prod_{i=1}^{I_k}\prod_{j=1}^{J_k}\phi^{i,j}(a_k^i, b_k^j), \nn
\end{align}
where\vspace{-2mm}
\begin{align}
	\phi^{i,j}(a_k^i, b_k^j)  =	\left\{
	\begin{array}{lr}
		0, &a_k^i = j, b_k^j\neq i\\
		&b_k^j = i, a_k^i\neq j\\
		1, &\text{otherwise}. \nn
	\end{array}
	\right.
\end{align}
The factor $\Phi(\V{a}_k,\V{b}_k) = 1$ if $\V{a}_k$ and $\V{b}_k$ describe the same valid data association event. For a legacy PO $i$, the state likelihood ratio is defined as
\begin{align}
\label{lik_legacy_PO}	q(\underline{\V{x}}_k^i,1, a_k^i; \V{z}_k) =&~\left\{
	\begin{array}{lr}
		\frac{p_df(\V{z}^j_k|\underline{\V{x}}_k^i)}{\mu_{\text{FA}}f_{\text{FA}}(\V{z}^j_k)}, &a_k^i = j \\
		1 - p_d, &a_k^i = 0
	\end{array}
	\right.\\[1mm]
	q(\underline{\V{x}}_k^i,0, a_k^i; \V{z}_k) =&~1(a_k^i) \nn
\end{align}
where $p_d$ denotes the detection probability and $1(a_k^i)$ denotes the indicator function, i.e. $1(a_k^i)=1$ if $a_k^i=0$ and $0$ otherwise. If a measurement $j\in\{1,\dots,J_k\}$ is generated by a PO $i$, it is distributed according to $f(\V{z}^j_k|\underline{\V{x}}_k^i)$. If the measurement $j$ is not generated by any PO, it is considered as the false alarm measurement, which is independent and identically distributed (i.i.d.) according to $f_{\text{FA}}(\cdot)$. The number of false alarm measurements is modeled by a Poisson distribution with mean $\mu_{\text{FA}}$. For new PO $j$, the state likelihood ratio is defined\vspace{0mm} as
\begin{align}
\label{lik_new_PO}	v(\overline{\V{x}}_k^j,1, b_k^j; \V{z}^j_k) =&~\left\{
	\begin{array}{lr}
		\frac{\mu_uf_u(\overline{\V{x}}_k^j)f(\V{z}^j_k|\overline{\V{x}}_k^j)}{\mu_{\text{FA}}f_{\text{FA}}(\V{z}^j_k)}, &b_k^j = 0\\
		0, &b_k^j \ne 0
	\end{array}
	\right.\\[1mm]
	v(\overline{\V{x}}_k^j,0, b_k^j; \V{z}^j_k) =&~f_{D}(\overline{\V{x}}_k^j).
\end{align}
New POs are i.i.d. according to $f_u(\overline{\V{x}}_k^j)$, and the number of new POs is modeled by a Poisson distribution with mean $\mu_u$. A detailed derivation of $v(\cdot)$ and $q(\cdot)$ is provided in \cite{LiaMey:J24, BP-2018Meyer}\vspace{0mm}.

\subsection{Object Declaration and State Estimation}

In our Bayesian setting, we declare the existence of POs and estimate their state based on the marginal existence probabilities $p(r^{i}_{k} \rmv=\rmv 1 | \V{z}_{1 : k})$ and the conditional PDF $f(\V{x}^{i}_{k,} | r^{i}_{k} \rmv\rmv=\rmv\rmv 1, \V{z}_{1 : k})$. More specifically, a PO is declared to exist if $p(r^{i}_{k} \rmv=\rmv 1 | \V{z}_{1 : k})$ is above the threshold  $T_{\text{dec}}$. For PO $i$ that is declared to exist, a state estimate of $\V{x}^{i}_{k}$ is then provided by the minimum-mean-square-error (MMSE) estimator \cite{Kay:B93}, i.e.,
\begin{equation}
 \hat{\V{x}}^{i}_{k} \rmv=\rmv \int \V{x}^{i}_{k} \ist f(\V{x}^{i}_{k}  \ist|\ist r^{i}_{k} \rmv=\rmv 1, \V{z}_{1 : k}) \ist \mathrm{d}\V{x}^{i}_{k}. \nn
 \vspace{0mm}
 \end{equation} 
Note that the existence probability and the conditional PDF are computed as $p(r^{i}_{k} \rmv= 1 \ist | \ist \V{z}_{1 : k}) \rmv=\rmv \int f(\V{x}^{i}_{k}, r^{i}_{k} \rmv=\rmv 1 \ist |$ $\V{z}_{1 : k}) \mathrm{d}\V{x}^{i}_{k}$ and $f(\V{x}^{i}_{k} \ist | \ist r^{i}_{k} \rmv=\rmv 1, \V{z}_{1 : k}) \rmv=\rmv f(\V{y}^{i}_{k} \ist | \ist \V{z}_{1 : k}) / p(r^{i}_{k} = 1 \ist | \ist \V{z}_{1 : k})$. Thus, both tasks require computation of the marginal posterior PDF $f(\V{y}^{i}_{k} \ist | \ist \V{z}_{1 : k}) \rmv\triangleq\rmv f(\V{x}^{i}_{k}, r^{i}_{k} \ist | \ist \V{z}_{1 : k})\vspace{0mm}$.  By applying BP following \cite[Sec. VIII-IX]{BP-2018Meyer}, accurate approximations (a.k.a. ``beliefs'') $\tilde{f}(\V{y}^{i}_{k}) \rmv\approx\rmv f(\V{y}^{i}_{k} | \V{z}_{1 : k})\vspace{-.5mm}$ of marginal posterior PDFs are obtained efficiently. 
 
Since a new PO is introduced for each measurement, the number of POs grows with time $k$. Thus, POs whose approximate existence probability is below a threshold $T_{\text{pru}}$ are removed from the state space\vspace{0mm}.

\section{The Methodology}
\vspace{-1mm}
This section introduces the proposed NEBP+ method. It first reviews the factor graph and BP for MOT and then introduces the new neural architecture.  
\begin{figure*}[!t]
	\centering
	\includegraphics[width=6.3in]{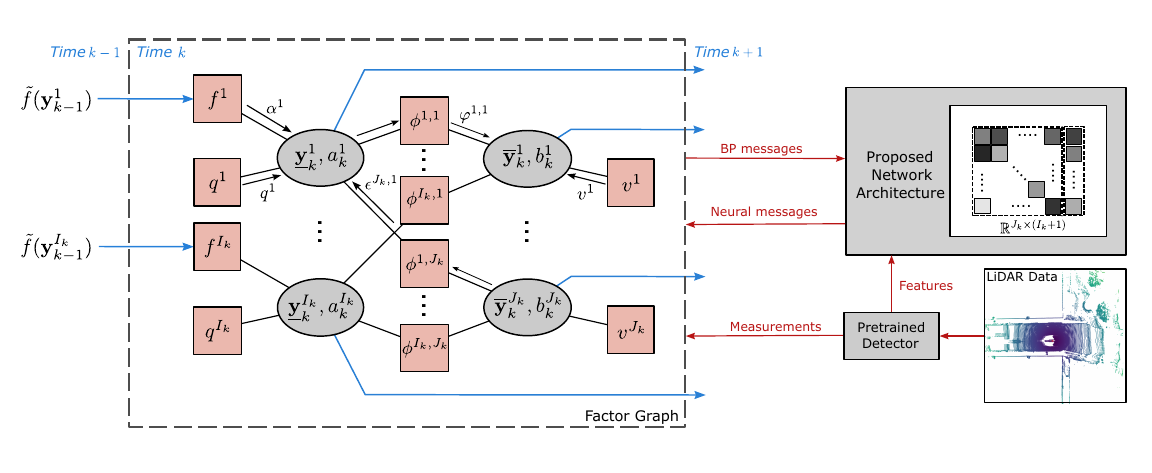}
	\vspace{-5mm}
	\caption{Block diagram of the proposed NEBP+ approach to MOT. The factor graph, BP messages, and message exchange between the factor graph and neural architecture are shown. The factor graph processes the measurements provided by the detector at the current time step and beliefs from the previous time step. The resulting BP messages are passed to the neural architecture. The neural architecture computes the neural messages based on BP messages and a variety of features provided by the pre-trained detector. (More details on the processing performed by the neural architecture are shown in Fig.~\ref{neural network}.) Finally, the neural messages are passed back to enhance the data association process and track initialization. The following shorthand notation is used: $f^i = f(\underline{\V{y}}^i_{k}|{\V{y}}^i_{k-1})$, $q^i = q(\underline{\V{y}}_k^i, a_k^i; \V{z}_k)$, $v^j = v(\overline{\V{y}}_k^j, b_k^j; \V{z}_k)$, $\phi^{i,j} = \phi^{i,j}(a_k^i, b_k^j)$, $\alpha^i =  \alpha_{k}(\underline{\V{x}}^i_{k},\underline{r}^i_{k})$, $\varphi^{i,j} = \varphi_k^{i,j,[\ell]}(b_k^j)$, and $\epsilon^{j,i} = \epsilon_k^{j,i,[\ell]}(a_k^i)$ \vspace{-2mm}.}
	\label{factor graph}
\end{figure*}
\subsection{Factor Graph and BP for MOT}
With the assumptions in section II and other common assumptions \cite{BP-2018Meyer}, the factorization of the joint posterior PDF $f(\V{y}_{0:k},\V{a}_{1:k},\V{b}_{1:k}|\V{z}_{1:k})$ is given as\vspace{1mm} follows
\begin{align}
	f&(\V{y}_{0:k},\V{a}_{1:k},\V{b}_{1:k}|\V{z}_{1:k}) \notag\\
	&\hspace{3mm}\propto \left(\prod_{n' = 1}^{N_0}f(\V{y}^{n'}_{0})\right)\prod_{k' = 1}^{k}\left(\prod_{n=1}^{N_{k'-1}}f(\underline{\V{y}}^n_{k'}|{\V{y}}^n_{k'-1})\right)\notag\\
	&\hspace{3mm}\times \left(\prod_{i=1}^{I_{k'}}q(\underline{\V{y}}^i_{k'},a^i_{k'};\V{z}_{k'})\prod_{j'=1}^{J_{k'}}\phi^{i,j'}(a^i_{k'}, b^{j'}_{k'})\right)\notag\\
	&\hspace{3mm}\times\prod_{j=1}^{J_{k'}}v(\overline{\V{y}}^j_{k'},b^j_{k'};\V{z}^j_{k'}).\nn \\[-3mm]
	\nn
\end{align} 
The factorization above provides the basis for a factor graph representation in Fig.~\ref{factor graph} \cite{LiaMey:J24}. Since the considered factor graph has loops, the order of message passing is given as (i) BP messages are only sent forward in time, (ii) iterative message passing is only performed for data association and at each time step individually.  BP for MOT consists of the following three steps.
\subsubsection{Prediction}
The prediction step uses the same principle as \cite{BP-2018Meyer}, where the messages passing from factor nodes $f(\underline{\V{y}}^i_{k}|\V{y}^i_{k-1})$ to the variable nodes $(\underline{\V{y}}^i_{k}, a_k^i)$ (See Fig.\ref{factor graph}) are computed as
\begin{align}
	\alpha_{k}(\underline{\V{x}}^i_{k},\underline{r}^i_{k}) =& \sum_{\underline{r}^i_{k-1} \in \{0,1\}} \int{f(\underline{\V{x}}^i_{k},\underline{r}^i_{k}|{\V{x}}^i_{k-1},{r}^i_{k-1})} \nn\\[1mm]
	&\times \tilde{f}({\V{x}}^i_{k-1},{r}^i_{k-1})d\V{x}^i_{k-1},\notag
\end{align} 
where $\tilde{f}(\cdot)$ denotes beliefs computed at the last time step \cite{BP-2018Meyer}. 
\subsubsection{Iterative Probabilistic Data Association}
After the prediction step, an iterative probabilistic data association step is performed for each legacy and new PO. In brief, the objective of this step is to find the global soft data association by exchanging information of local likelihood evaluation among all nodes $(\underline{\V{y}}_k^i, a_k^i)$ and $(\overline{\V{y}}_k^j, b_k^j)$.
\par At message passing iteration $\ell \in\{1,2,\dots,L\}$, the messages $\varphi_k^{i,j,[\ell]}(b_k^j)$ and $\epsilon_k^{j,i,[\ell]}(a_k^i)$ pass from the indicator factor $\phi^{i,j}(a_k^i,b_k^j)$ to the variable node $(\overline{\V{y}}_k^j, b_k^j)$ and $(\underline{\V{y}}_k^i, a_k^i)$. The entire iterative steps are given as
\begin{align}
\varphi_k^{i,j,[\ell]}(b_k^j) & = \sum_{a_k^i=0}^{J_k}\beta_k^i(a_k^i)\phi^{i,j}(a_k^i,b_k^j)\prod_{ j' = 1\atop j' \neq j}^{J_k}\epsilon_k^{j',i,[\ell]}(a_k^i)\\
\epsilon_k^{j,i,[\ell]}(a_k^i) & =\sum_{b_k^j=0}^{I_k}\xi^j_k(b^j_k)\phi^{i,j}(a_k^i,b_k^j)\!\prod_{ i' = 1\atop i' \neq i }^{I_k}\varphi_k^{i',j,[\ell-1]}(b_k^j)\,
\end{align}
where
\begin{align}
	\label{DA_beta}\beta^{i}_{k}(a^i_k) = & \sum_{\underline{r}_k^i = \{0,1\}}\int{q(\underline{\V{x}}_k^i, \underline{r}_k^i, a_k^i; \V{z}_k)\alpha_k(\underline{\V{x}}_k^i, \underline{r}_k^i)d\underline{\V{x}}_k^i}\\
	\label{DA_xi} \xi ^{j}_{k}(b^j_k) = & \sum_{\overline{r}_k^j = \{0,1\}}\int{v(\overline{\V{x}}_k^j,\overline{r}_k^j, b_k^j; \V{z}^j_k)d\overline{\V{x}}_k^j},
\end{align}
The iterative message passing process is initialized by setting $\varphi_k^{i,j,[0]}(b_k^j) = 1$ for all $j \in \{1,2,\dots,J_k\}$ and $i \in \{1,2,\dots,I_k\}$. 
\begin{figure*}[!t]
	\centering
	\includegraphics[scale=0.68]{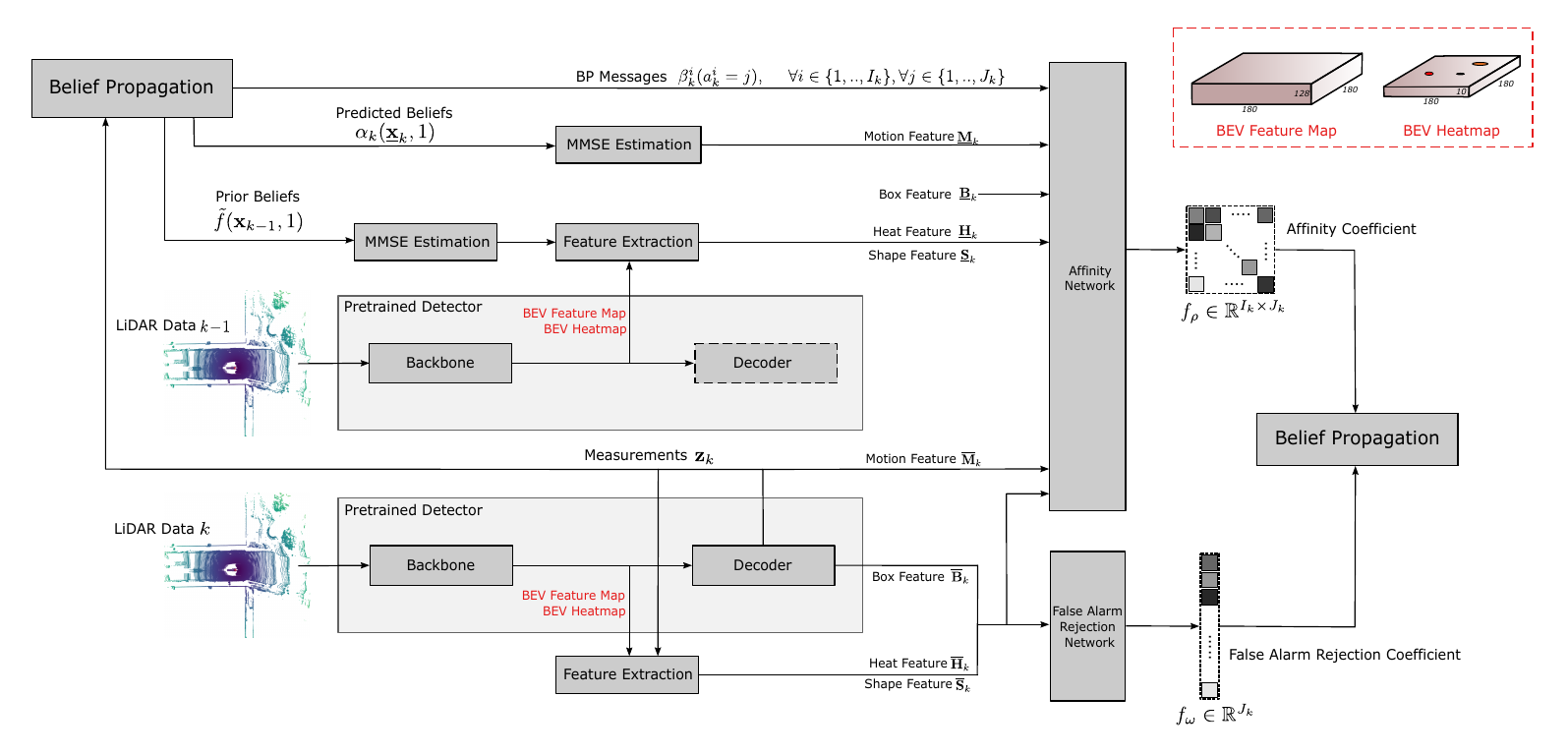}
	\vspace{-7.5mm}
	\caption{Neural architecture of the proposed NEBP+ approach.  First, motion, box, shape, and heat features are extracted for each measurement and each PO. Next, an affinity coefficient is computed for each pair of PO and measurement, and a false alarm rejection coefficient is computed for each measurement. As discussed in Section~\ref{sec:neuNet}, the affinity coefficients are computed based on a linear combination of feature similarity matrixes and BP messages. The false alarm rejection coefficients are calculated based on the box, shape, and heat features extracted for each measurement by the detector\vspace{-2mm}.}
	\label{neural network}
\end{figure*}
\subsubsection{Belief Update}
After the last message passing iteration $\ell = L$, the beliefs $\tilde{f}(\underline{\V{y}}^i_k, a^i_k),~i\in\{1,2,\dots,I_k\}$ and $\tilde{f}(\overline{\V{y}}^j_k, b^j_k),~j\in\{1,2,\dots,J_k\}$ are computed according to
\begin{align}
	\tilde{f}(\underline{\V{y}}^i_k, a^i_k) =&~ \frac{1}{\underline{C}_{k}^i}q(\underline{\V{y}}^i_k,a^i_k;\V{z}_k)\alpha_k(\underline{\V{y}}^i_k)\prod_{j=1}^{J_k}\epsilon^{j,i,[L]}(a_k^i) \nn\\[1mm]
	\tilde{f}(\overline{\V{y}}^j_k, b^j_k) =&~ \frac{1}{\overline{C}_{k}^j}v(\overline{\V{y}}^j_k,b^j_k;\V{z}^j_k)\prod_{i=1}^{I_k}\varphi^{i,j,[L]}(b_k^j) \nn
\end{align}
where both notations $\underline{C}_{k}^i, \overline{C}_{k}^j$ are normalized constants to make sure that the final beliefs $\tilde{f}(\underline{\V{y}}^i_k, a^i_k), \tilde{f}(\overline{\V{y}}^j_k, b^j_k)$ integrate to unity. By performing marginalization for them to obtain $\tilde{f}(\underline{\V{y}}^i_k), \tilde{f}(\overline{\V{y}}^j_k)$, these beliefs are used for object declaration and estimation \cite{BP-2018Meyer} and propagated to the next time\vspace{0mm} step.

\subsection{The Neural Architecture\vspace{-.5mm}}
\label{sec:neuNet}

In this section, we will discuss how the proposed neural architecture extracts features from LiDAR data and computes two types of coefficients based on the extracted features. The coefficients are used to enhance data association and object initialization performed by BP.
\subsubsection{Feature Extraction}
\par For each legacy PO $i$ at time $k$, the motion feature  $\underline{\M{M}}^i_k \in \mathbb{R}^4$ is obtained by MMSE estimation. The motion feature consists of the objects' 2D position and velocity. The box feature $\underline{\M{B}}^i_k \in \mathbb{R}^3$ consists of the width, height, and length of the bounding box that indicates the size of a PO. The box feature is extracted together with measurements that initiated the PO.

\par To extract the shape and heat features of POs, the raw data (LiDAR point cloud) is passed into the pre-trained backbone \cite{Voxelnet-2018Zhou} of the FocalFormer3D detector \cite{Focalformer3d-2023chen}. The pretrained backbone provides a bird’s-eye-view (BEV) \cite{BEVReview-2023Li} feature map and heatmap, a unified representation of the LiDAR data in the coordinate system of the sensing vehicle. The BEV feature map is generally a 3D tensor map, where each region of the map encodes a high dimensional volumetric representation of the object shape information in that region. The BEV heatmap is also a 3D tensor map, which includes the presence and category information of potential objects throughout the entire map \cite{Voxelnet2-2018Yan}. Then, we extract the shape $\underline{\M{S}}^i_k \in \mathbb{R}^{64}$ and heat $\underline{\M{H}}^i_k \in \mathbb{R}^{32}$ features from these two maps respectively through an encoder network (See Fig. \ref{neural network}). The process discussed above can be summarized as 
\begin{align}
	(\underline{\M{S}}^i_k, \underline{\M{H}}^i_k) = {\cal{D}}({\Set{Z}}_{k-1};{\widehat{\V{x}}}_{k-1}^i), 
	\label{eq:D1}
\end{align}
where ${\widehat{\V{x}}}_{k-1}^i$ is the MMSE estimation for a legacy PO $i$ at time $k-1$, ${\Set{Z}}_{k-1}$ denotes the raw data at time $k-1$. The network ${\cal{D}}$ consists of the pre-trained backbone of the detector and an additional encoder network. For each measurement indexed by $j$, the motion feature $\overline{\M{M}}^j_k \in \mathbb{R}^4$ is equal to the measurement  $\V{z}_k^j$ itself while the box feature $\overline{\M{B}}^j_k \in \mathbb{R}^3$ is also provided by the detector. Similarly, the shape and heat features of each measurement $j$ are obtained\vspace{.5mm} as
\begin{align}
	(\overline{\M{S}}^j_k, \overline{\M{H}}^j_k) = {\cal{D}}({\Set{Z}}_{k};\V{z}_k^j)
	\label{eq:D2}
\end{align}
where $\overline{\M{S}}^j_k, \overline{\M{H}}^j_k$ share the same dimension with $\underline{\M{S}}^i_k, \underline{\M{H}}^i_k$, respectively. The 2D position information within ${\widehat{\V{x}}}_{k-1}^i$ and $\V{z}_k^j$ is used to select the corresponding region in the BEV feature map and heatmap. 

\subsubsection{The Affinity Coefficient}
By utilizing low-dimensional features (i.e., object position, motion, and size information) and high-dimensional features (i.e., features extracted from the raw LiDAR data), the proposed neural architecture is expected to obtain improved similarity information between objects and measurements. Similarity is represented by ``affinity coefficients'' that are used to improve data association performed by BP. In particular, the affinity coefficients enhance the association probabilities by taking the information provided by the statistical model, the measurements, and the raw sensor data into account.
\par Specifically, the affinity coefficient is computed through the ``Affinity Network'' (see Fig.~\ref{neural network}). By comparing the features of each legacy PO and measurement, the network computes four similarity matrices, the elements of which represent their similarity in the motion, box, shape, and heat features, respectively. For each PO $i$ and measurement $j$, the processing steps of the ``Affinity Network'' are as follows:
\begin{enumerate}		
	\item[-] The \textbf{motion feature} similarity is obtained as
	\begin{align}
		\V{L}_m^{i,j} =&~\underline{\M{M}}^i_k - \overline{\M{M}}^j_k \\[.6mm]
		R_m^{i,j} =&~{\cal{D}}_m(\V{L}_m^{i,j})
	\end{align}
	where $\V{L}_m^{i,j} \in \mathbb{R}^{4}$, $R_m\in \mathbb{R}$ and ${\cal{D}}_m$ is a neural network.
	\item[-] The \textbf{box feature} similarity is given by
	\begin{align}
		\V{L}_b^{i,j} =&~\underline{\M{B}}^i_k - \overline{\M{B}}^j_k \\[.6mm]
		R_b^{i,j} =&~{\cal{D}}_b(\V{L}_b^{i,j}) 
	\end{align}
	where $\V{L}_b^{i,j} \in \mathbb{R}^{3}$, $R_b\in \mathbb{R}$, ${\cal{D}}_b$ is a neural network.
	\item[-] The \textbf{shape and heat feature} similarities are obtained similarly, i.e.\vspace{-3mm},
	\begin{align}
		\V{L}_s^{i,j} =&~\underline{\M{S}}^i_k - \overline{\M{S}}^j_k \nn\\
		\V{L}_h^{i,j} =&~\underline{\M{H}}^i_k - \overline{\M{H}}^j_k \nn\\
		R_s^{i,j} =&~{\cal{D}}_s(\V{L}_s^{i,j}) \nn\\
		R_h^{i,j} =&~{\cal{D}}_h(\V{L}_h^{i,j})\nn
	\end{align}
	where $R_s^{i,j}, R_h^{i,j} \in \mathbb{R}$ represent the shape and heat feature similarities respectively, and ${\cal{D}}_s, {\cal{D}}_h $ are neural networks. 
	\item[-] The \textbf{overall similarity matrix} is a weighted sum of the four matrices discussed above and the BP messages. These weights are also provided by a neural network, where we concatenate all distance $\V{L}_m^{i,j}, \V{L}_b^{i,j}, \V{L}_s^{i,j}$, and $\V{L}_h^{i,j}$ calculated above. The weight output $\V{\omega}^{i,j}$ is computed as
	\begin{align}
		\V{\Omega}^{i,j} =&~ [(\V{L}_m^{i,j})^\top~(\V{L}_b^{i,j})^\top~(\V{L}_s^{i,j})^\top~(\V{L}_h^{i,j})^\top]^\top \nn \\
		\V{\omega}^{i,j} =&~ {\cal D}_w(\V{\Omega}^{i,j}) \nn
	\end{align}
	where $\V{\omega}^{i,j} \in {\left(0,1\right)}^{5}$ and $\V{\Omega}^{i,j}\in\mathbb{R}^{103}$. The neural network ${\cal D}_w$ is expected to capture the quality of different features to dynamically assign weights to different types of similarities. Then, taking each element in $\V{\omega}^{i,j}$ as the corresponding weight for different types of features, the overall pairwise similarity $R^{i,j}$ is obtained by a linear combination, i.e.\vspace{-.5mm}, 
	\begin{align} \nn
		R^{i,j} =(\V{\omega}^{i,j})^\top \V{R}^{i,j}
	\end{align}
	where $\V{R}^{i,j} = [R_m^{i,j}, R_b^{i,j}, R_s^{i,j}, R_h^{i,j}, R_{\text{BP}}^{i,j}]^\top \in\mathbb{R}^5$ and $R_{\text{BP}}^{i,j} =  \beta^{i}_{k}(a^i_k = j)$ denotes the BP message in \eqref{DA_beta}. $R_{\text{BP}}^{i,j}$ can be seen as the similarity information provided by the statistical model.
	\end{enumerate}
\par After obtaining the overall similarity matrix, the last step is to apply a learnable nonlinear transformation to it, i.e., we use a neural network ${\cal D}_{\text{Aff}}$ to further process $R^{i,j}$ 
	\begin{align}
		f^{i,j}_\rho =&~{\cal D}_{\text{Aff}}(R^{i,j}) \nn
	\end{align}
where $f^{i,j}_\rho \in \mathbb{R}$ denotes the affinity coefficient, which includes the similarity information between a specific PO $i$ and a measurement $j$.
\subsubsection{The False Alarm Rejection Coefficient}
To decrease the influence brought by false alarms, the ``False Alarm Rejection Network'' ${\cal D}_{\text{Far}}$ is designed to identify which measurements are likely false alarms (see Fig.~\ref{neural network}). Then, for the measurement identified as a potential false alarm, the false alarm distribution in the statistical model used by BP will be locally increased. This false alarm rejection coefficient will help us reduce the probability that the measurement is associated with an existing object and initialize a new object track. The false alarm rejection coefficient $f^j_{\omega}$ for each measurement $j$ is computed\vspace{0mm} by
\begin{align}
	f^j_\omega =&~{\cal D}_{\text{Far}}(\overline{\M{B}}^j_k, \overline{\M{S}}^j_k, \overline{\M{H}}^j_k), \nn \\[-5mm]
	\nonumber
\end{align}
where $f^j_\omega \in (0,1)$ and ${\cal D}_{\text{Far}}$ is an MLP with a sigmoid function in the output layer\vspace{-.5mm}.
 
\subsection{Neural Enhanced BP}

The neural messages $f^j_\omega,f^{i,j}_\rho$ are used to enhance the likelihood ratios $q(\underline{\V{x}}_k^i,\underline{r}_k^i,a_k^i;\V{z}_k)$ and $v(\overline{\V{x}}_k^j,\overline{r}_k^j, b_k^j;\V{z}^j_k)$. As in \cite{LiaMey:J24}, we calculated the normalized versions of the original BP messages, i.e.,
\begin{align}
	q_s&(\underline{\V{x}}_k^i,\underline{r}_k^i,a_k^i;\V{z}_k) \notag\\
	&~~~~~=\frac{q(\underline{\V{x}}_k^i,\underline{r}_k^i,a_k^i;\V{z}_k)}{\sum_{a_k^i = 0}^{J_k}\sum_{\underline{r}_k^i = \{0,1\}}\int{q(\underline{\V{x}}_k^i,\underline{r}_k^i,a_k^i;\V{z}_k)d\underline{\V{x}}_k^i}}, \nn \\[-6mm]
	\nn
\end{align}
\begin{align}
	v_s&(\overline{\V{x}}_k^j,\overline{r}_k^j, b_k^j;\V{z}^j_k)  \notag\\
	&~~~~~=\frac{v(\overline{\V{x}}_k^j,\overline{r}_k^j, b_k^j;\V{z}^j_k)}{\sum_{b_k^j = 0}^{I_k}\sum_{\overline{r}_k^j = \{0,1\}}\int{v(\overline{\V{x}}_k^j,\overline{r}_k^j,b_k^j;\V{z}^j_k)d\overline{\V{x}}_k^j}}. \nn
\end{align}
Finally, the normalized messages are enhanced\vspace{2mm} as
\begin{align}
	\widehat{q}_s(\underline{\V{x}}_k^i,1,a_k^i=j;\V{z}_k) =&~ C^{i,j} \cdot {q}_s(\underline{\V{x}}_k^i,1,a_k^i=j;\V{z}_k), \nn\\[2mm]
	\widehat{v}_s(\overline{\V{x}}_k^j,1, b_k^j=0;\V{z}^j_k) =&~ f^j_\omega\cdot v_s(\overline{\V{x}}_k^j,1,b_k^j=0;\V{z}^j_k), \nn\\[-4mm]
	\nn
\end{align}
where we\vspace{.5mm} used
\begin{align}
		C^{i,j} =&~ f^j_\omega +  \frac{f_{\text{relu}}(f^{i,j}_\rho)}{\hat\beta^{i,j}},\nn\\[3mm]
		\hat\beta^{i,j} =&~ \frac{\beta_k^{i}(a^i_k = j)}{\sum_{j=0}^{J_k}\beta_k^{i}(a^i_k = j)}.\nn \\[-5mm]
		\nn
\end{align}
Here, $f_{\text{relu}}(\cdot)$ denotes  rectified linear unit. In other words, the BP messages passed from the likelihood ratio node are enhanced using the affinity and false alarm rejection coefficients. This enhancement contributes to refining the iterative data association process by recomputing messages $\beta^{i}_{k}(a_k^i)$ and $\xi^{j}_{k}(b_k^j)$ in \eqref{DA_beta}, \eqref{DA_xi} using $\widehat{q}_s(\cdot)$ and $\widehat{v}_s(\cdot)$ respectively.

\subsection{The Loss Function}
\vspace{-.5mm}
We use a ``weighted'' binary cross-entropy loss for the loss function calculation. First, for the affinity coefficient $f_{\rho}$, the corresponding loss ${\cal L}_{A}$ is calculated\vspace{1mm} by
\begin{align}
\label{L_A1} {\cal L}_{A1} =&~ - \frac{\sum_{i=1}^{I_k}\sum_{j=1}^{J_k} f^{i,j}_{gt,\rho}\ln(f_\sigma(f^{i,j}_\rho))}{\sum_{i=1}^{I_k}\sum_{j=1}^{J_k} f^{i,j}_{gt,\rho}}\\[3mm]
\label{L_A2} {\cal L}_{A2} =&~-\frac{\sum_{i=1}^{I_k}\sum_{j=1}^{J_k}(1 - f^{i,j}_{gt,\rho})\ln(1 - f_\sigma(f^{i,j}_\rho))}{\sum_{i=1}^{I_k}\sum_{j=1}^{J_k} (1-f^{i,j}_{gt,\rho})}\\[2mm]
	{\cal L}_{A} =&~{\cal L}_{A1} + {\cal L}_{A2},
\end{align}
where $f_\sigma$ denotes the sigmoid function and $f^{i,j}_{gt,\rho}$ denotes the ground truth data association result, i.e., if the PO $i$ is indeed matched with measurement $j$, then $f^{i,j}_{gt,\rho}=1$ and otherwise $f^{i,j}_{gt,\rho}=0$. Please refer to \cite{LiaMey:J24} for detailed steps on obtaining the ground truth. ${\cal L}_{A1}$ represents the accuracy of estimating the correct data association, while ${\cal L}_{A2}$ represents the accuracy of estimating no association. The reason for using the normalization term in \eqref{L_A1} and \eqref{L_A2} is to deal with the data imbalance problem, i.e., it is equally important whether a PO $i$ is associated with a measurement $j$ or not, regardless of the number of POs and measurements. Compared to \cite{LiaMey:J24}, the proposed loss function emphasizes the significance of correct measurement-to-objects associations.
\par For the false alarm rejection coefficient $f_{\omega}$, the corresponding loss function is given\vspace{1mm} as
\begin{align}
	{\cal L}_{F1} =&~-\frac{\sum_{j=1}^{J_k} f^j_{gt,\omega}\ln(f^j_\omega)}{\sum_{j=1}^{J_k} f^j_{gt,\omega}} \\[1.5mm]
	{\cal L}_{F2} =&~ -u \cdot\frac{\sum_{j=1}^{J_k}(1 - f^j_{gt,\omega})\ln(1 - f^j_\omega)}{\sum_{j=1}^{J_k} (1-f^j_{gt,\omega})} \\[1.5mm]
	\label{L_A3}  {\cal L}_{F} =&~{\cal L}_{F1} + {\cal L}_{F2}
\end{align}
where $f^j_{gt,\omega}$ represents the ground truth label for each measurement \cite{LiaMey:J24} and $u\in[0,1]$ is a tuning parameter. ${\cal L}_{F1}$ represents our classification accuracy for true measurements. ${\cal L}_{F2}$ represents our classification accuracy for false alarms. The tuning parameter is introduced because missing an object is usually more harmful than having a false\vspace{0mm} alarm.

\section{Numerical Analysis}

The section presents results of nuScenes LiDAR autonomous driving dataset \cite{Nuscenes} to validate our method. This dataset consists of 1000 autonomous driving scenes and seven object classes. We use the official split of the dataset, where there are 700 scenes for training, 150 for validation, and 150 for testing. Each scene lasts roughly 20 seconds and contains keyframes sampled at 2Hz. The recently introduced FocalFormer3D detector provides measurements for MOT \cite{Focalformer3d-2023chen} that have been pre-processed using the non-maximum suppression (NMS) technique \cite{NMS-2006Neubeck}. We use pre-trained versions of the FocalFormer3D detector and its\vspace{0mm} backbone.

\subsection{Implementation Details}
We use the particle-based implementation of BP-based MOT, where the number of particles is set to $10^4$. The state of a PO $\V{x}_k^i \in \mathbb{R}^{6}$ consists of its 2D position, velocity, and acceleration. A constant-acceleration motion model models the object dynamics. The measurement $\V{z}_k^j \rmv\in\rmv \mathbb{R}^{4}$ consists of the 2D position and velocity. Measurements are modeled using a linear Gaussian measurement model. The corresponding likelihood ratio is used for compution of $q(\underline{\V{x}}_k^i,1, a_k^i; \V{z}_k)$ and $v(\overline{\V{x}}_k^j,1, b_k^j; \V{z}^j_k)$ in \eqref{lik_legacy_PO} and \eqref{lik_new_PO}. We also use 3D bounding box information and the detection score $s_{k}^j \in (0,1]$ provided by the pre-trained detector. In particular, for each new PO, the corresponding bounding box is stored and used as a box feature in the affinity network (see Fig.~\ref{neural network}). Together with the probability of existence, the detection score is used to compute the final AMOTA score for each PO (see \cite{LiaMey:J24} for details).

\par The region of interest (ROI) is given by $[P_x - 54, P_x + 54] \times [P_y - 54, P_y + 54] $, where $P_x, P_y$ represent the 2D positions of the ego vehicle. The prior PDF of false alarms $f_{\text{FA}}(\cdot)$ and newly detected objects $f_u(\cdot)$ are uniformly distributed over the ROI. All other parameters used in the proposed NEBP+ method are estimated from the training data as in \cite{LiaMey:J24} or set as in \cite{LiaMey:J24}.
\par The encoder network within the neural architecture ${\cal{D}}$ (cf.~\eqref{eq:D1} and \eqref{eq:D2}) consists of two convolution layers followed by an MLP, which is used to extract heat and shape features. The neural networks ${\cal{D}}_m, {\cal{D}}_b,{\cal{D}}_s,{\cal{D}}_h, {\cal{D}}_{\omega}, {\cal{D}}_{\text{Aff}}, {\cal{D}}_{\text{Far}}$ are all MLPs with leaky ReLU activation functions in the hidden layers\vspace{1mm}.

\subsection{Performance Evaluation}
\subsubsection{Overall Performance}
Two primary metrics for evaluating nuScenes are Average Multi-Object Tracking Accuracy (AMOTA) and Average Multi-Object Tracking Precision (AMOTP) \cite{Nuscenes}. AMOTA includes errors such as false alarms, missed objects, and identity switches, while AMOTP consists of position errors between ground truth and estimated tracks. Details about these metrics are provided in \cite{Nuscenes}.
\begin{table}[!t]
	\centering
	\caption{Performance of Nuscenes Test Set}
	\label{test set}
	\resizebox{\linewidth}{!}
	{\begin{tabular}{c|c|c|c} 
		\hline
		Method& Modalities &\textbf{AMOTA$\uparrow$} &AMOTP$\downarrow$\\ \hline
		\hline
		Poly-MOT\cite{PolyMOT-2023Li}&LiDAR+Camera&75.4&42.2\\   
		CAMO-MOT\cite{CamoMOT-2023Wang}&LiDAR+Camera&75.3&47.2\\ 
		\hline  
		\rowcolor{green!25}\textbf{NEBP+}&LiDAR&74.6&49.8\\ 
		\hline  
		BEVFusion\cite{BEVFusion-2023Liu}&LiDAR+Camera&74.1&40.2\\ 
		MSMDFusion\cite{MSMDFusion-2023Jiao}&LiDAR+Camera&74.0&54.9\\ 
		FocalFormer3D-F\cite{Focalformer3d-2023chen}&LiDAR+Camera&73.9&51.4\\ 
		3DMOTFormer\cite{FormerMOT3D-2023Shuxiao} &LiDAR+Camera&72.5&53.9\\ 
		FocalFormer3D\cite{Focalformer3d-2023chen}&LiDAR&71.5&54.9\\ 
		VoxelNeXt\cite{VoxelNext-2023Chen}&LiDAR&71.0&51.1\\\hline
	\end{tabular}
}
\vspace*{-3mm}
\end{table}

\par The performance of the proposed NEBP+ method applied to nuScenes test data is shown in Tables \ref{test set}. It outperforms all LiDAR-only reference methods in terms of both AMOTA and AMOTP performance. The improved performance of NEBP+ can be explained by the fact that it combines the well-established statistical model for MOT \cite{JPDA-1998,MHT-2004Blackman,BP-2018Meyer} with learned information provided by the neural architecture.

\subsubsection{Ablation Study}
Next, we present results based on the nuScenes validation data. We compare the performance of (i) traditional BP-based MOT, (ii) NEBP proposed in \cite{LiaMey:J24} that makes use of the loss functions in \eqref{L_A1}--\eqref{L_A3}, (iii) NEBP+ that only uses motion features and BP messages as the input of the affinity network, (iv) NEBP+ that only uses shape and heat feature as the input of the affinity network. (v) NEBP+ that only uses the affinity coefficient, and (vi) NEBP+ that only uses the false alarm rejection coefficient. AMOTA and AMOTP results are shown in Table \ref{val set}.
\begin{table}[!t]
	\centering
	\caption{Performance of Nuscenes Validation Set}
	\label{val set}
	\resizebox{\linewidth}{!}{
		\begin{tabular}{c|c|c} 
			\hline
			Method &{\textbf{AMOTA}$\uparrow$} &AMOTP$\downarrow$\\ \hline
			\hline
			BP\cite{BP-2018Meyer}&73.3&50.8\\   
			NEBP \cite{LiaMey:J24}&74.3&54.8\\ 
			NEBP+ (Only Affinity Coeff. with Motion and BP Feat.)&73.9&51.5\\ 
			NEBP+ (Only Affinity Coeff. with Shape and BP Feat.)&74.2&51.9\\ 
			NEBP+ (Only Affinity Coeff.)&75.1&54.3\\ 
			NEBP+ (Only False Alarm Rejection Coeff.)&74.2&55.0\\ 
			\hline
			\rowcolor{green!25}\textbf{NEBP+ (Proposed)}&75.3&50.9\\\hline
		\end{tabular}}
		\vspace{-3mm}
\end{table}
\par We have the following observations based on Table \ref{val set}. (i) The proposed NEBP+ approach yields the largest improvement in AMOTA performance compared to BP. (ii) While NEBP \cite{LiaMey:J24} has improved AMOTA performance compared to BP, this comes at the cost of a reduced AMOTP performance. NEBP+, however, has significantly improved AMOTA performance compared to BP while yielding essentially identical AMOTP performance. (iii) In the considered scenario, performance improvements related to the affinity coefficient are much greater than those related to the false alarm rejection coefficient. (iv) Although using only a subset of features increases tracking performance compared to the traditional BP, the best result is achieved when all types of information are leveraged. (v) Even if only low-dimensional information (i.e., motion features and BP messages) is used as input for neural networks, this leads to improved tracking performance. Observation (v) can have two potential explanations. The Gaussian likelihood function and corresponding measurement variance used by model-based BP may not accurately describe the underlying data-generating process. The neural architecture may also have captured complex relationships across motion features that cannot be described by the traditional MOT model used by BP\vspace{-3mm}.

\section{Conclusion}
\vspace{-.5mm}
In this paper, we introduce NEBP+ for MOT. NEBP+ enhances both data association and track initialization of model-based BP by leveraging information learned from raw LiDAR data. Contrary to the original NEBP for MOT method, NEBP+ is based on an improved neural architecture that uses precomputed differences of features as input to the neural architecture and fuses different feature similarities based on learnable weights.  Evaluation results based on the nuScenes autonomous driving dataset demonstrate the state-of-the-art performance of NEBP+. The significant performance improvements of NEBP+ in the nuScenes autonomous driving challenge emphasize the superiority of combining well-established statistical models and neural architectures. As the framework of factor graphs and BP, NEBP+ is very flexible and can thus potentially be extended to a variety of 
further applications ranging from multipath-aided SLAM \cite{LeiMeyHla:J19,LeiVenTea:J23} to marine mammal tracking \cite{JanMeySnyWigBauHil:A22,ZhaMey:J24} 
Another venue for future research is the development of neural-enhanced track-before-detect methods \cite{LiaKroMey:J23}\vspace{1.5mm}.

\section{Acknowledgement}
\vspace{-.5mm}
This work was supported by the National Science Foundation (NSF) under CAREER Award No.~2146261. 
\vspace{3mm}

\pagebreak

\renewcommand{\baselinestretch}{1.1}
\selectfont

\bibliographystyle{IEEEtran}
\bibliography{References}

\end{document}